\documentclass[revtex4]{emulateapj}

\usepackage{graphicx}
\usepackage{hyperref}

\begin{document}

\title{Glitch or Anti-Glitch: a Bayesian View}%

\author{Yi-Ming Hu, Matthew Pitkin, Ik Siong Heng, Martin A. Hendry}%
\email[Yi-Ming Hu: ]{y.hu.1@research.gla.ac.uk}
\affiliation{SUPA, School of Physics and Astronomy, University of Glasgow}

\begin{abstract}
The sudden spin-down in the rotation of magnetar 1E 2259+586 observed by \citet{Archibald2013} was a rare event.
However this particular event, referred to as an  \emph{anti-glitch}, was followed by another event which \citet{Archibald2013} suggested could either be a conventional glitch or another anti-glitch.
Although there is no accompanied radiation activity or pulse profile change, there is decisive evidence for the existence of the second timing event, judging from the timing data. 
We apply Bayesian Model Selection to quantitatively determine which of these possibilities better explains the observed data.
We show that the observed data strongly supports the presence of two successive anti-glitches with a Bayes Factor, often called the odds ratio, greater than 40.
Furthermore, we show that the second anti-gtlich has an associated frequency change $\Delta\nu$ of $-8.2 \times 10^{-8}$ Hz.
We discuss the implications of these results for possible physical mechanisms behind this anti-glitch.
\end{abstract}

\maketitle
\section{Introduction}
Recently, \citet{Archibald2013} discovered an unexpected anti-glitch phenomenon in magnetar 1E 2259+586. Unlike a normal glitch, which undergoes  a sudden spin up, this magnetar experienced a sudden spin-down. The mechanism which caused this phenomenon is still under discussion \citep[e.g.][]{Tong2013a,Lyutikov2013,Katz2013,Huang2013,Ouyed2013}, but to our knowledge no model explicitly predicted an anti-glitch prior to this discovery, although \cite{Thompson2000} predict a similar potential phenomenon in SGR 1900+14.

The data analysis performed by \citet{Archibald2013} shows that during the observation, 1E 2259+586 undergoes two timing events separated by 50--90 days. The first event is a certain anti-glitch, while the nature of the second event is less certain. If it is also an anti-glitch this might require a qualitatively different physical model to explain its origin. Importantly, however, the analysis performed in \cite{Archibald2013} was unable to distinguish between these two types of glitch for the second event.

Since we know very little about the mechanism behind such a rare phenomenon, any information about it could be helpful to understand its physical cause. In this paper we seek to use the data themselves, employing the methods of Bayesian model selection, to distinguish between two competing models, wherein the second event is a glitch, or anti-glitch, respectively. More specifically, we compute the ratio of the \emph{evidence} for each model (as defined in Section \ref{sec:BI}) and investigate whether this ratio favours one model over the other.

The structure of this paper is as follows.  In Section \ref{sec:TM} we briefly review relevant details of the model for the time of arrival of pulses from the progenitor. Section \ref{sec:BI} then describes our Bayesian inference method for carrying out model selection.  Section \ref{sec:result} presents the results of our analysis, including a careful check on their robustness. Finally Section \ref{sec:CD} summarises our conclusions.

\section{Timing Model}\label{sec:TM}
The magnetar 1E 2259+586 was routinely observed by the X-ray Telescope (XRT) onboard Swift every 2-3 weeks, with more frequent observations being made shortly after discovering the first anti-glitch event reported in~\citet{Archibald2013}. The observations give the time of arrival (TOA) of each X-ray pulse (which can be corrected to the solar system barycenter), which in turn gives the pulse phase of the magnetar. Together with each TOA, the X-ray flux is also recorded. An increase in the X-ray flux helps to pinpoint the epoch of the first glitch event, while for the second event, no obvious flux change was detected -- thus contributing to the confusion about the second event's type.

We model the magnetar's phase evolution, $\phi(t)$, with the standard Taylor expansion of frequency and frequency derivatives \citep[e.g.][]{handbook} using terms up to second order. The effect of a (anti-)glitch on the phase timing model will be a sudden change in the frequency and frequency evolution after the event \citep[e.g. equation 11 of][]{Hobbs2006}. Therefore the difference between the phase model with and without the (anti-)glitch \citep[under the assumption that no decaying frequency increment is required due to none being fitted by][]{Archibald2013} is
\begin{equation}
\label{eq:ph_der}
\Delta \Phi_g(t) = \Delta\phi_g+\Delta\nu(t-T_g) + \frac{1}{2}\Delta\dot{\nu}(t-T_g)^2,
\end{equation}
where $T_g$ is the time of the (anti-)glitch, $\Delta\phi_g$ is an initial sudden phase change, and $\Delta\nu$ and $\Delta\dot{\nu}$ are respectively the difference in frequency and its first derivative before and after the glitch.

We further define $$R_i=\frac{\phi_i-N_i}{\nu}$$ as the time residual after subtracting the model predictions from the data. Here $\phi_i$ is the predicted pulse phase at the $i^{th}$ observation time $t_i$ (i.e.\ $\phi(t_i)$), $N_i$ is the exact phase at the TOA, which by definition is an integer, and $\nu$ is the frequency according to the model. Together with the observed timing uncertainty, $\sigma_i$, we can form
\begin{equation}\label{eq:res}
\chi^2 = \sum_{i=1}^{N}\left(\frac{R_i}{\sigma_i}\right)^2.
\end{equation}
In the next Section we will define the likelihood as proportional to $\exp(-\chi^2/2)$, and use it to evaluate the evidence for each model.

\section{Bayesian Inference}\label{sec:BI}

In Bayesian Inference, given prior background information ${\cal I}$ and observed data $D$, the posterior probability, $p(\theta|D,{\cal M},{\cal I})$, for a certain theoretical parameter set $\theta$ describing a model ${\cal M}$, follows from Bayes' theorem
\begin{equation}\label{bayes}
        p(\theta|D,{\cal M}, {\cal I})=\frac{p(D|\theta, {\cal M}, {\cal I})p(\theta|{\cal M},{\cal I})}{p(D|{\cal M},{\cal I})} .
\end{equation}
Here $p(D|\theta,{\cal M}, {\cal I})$ is the $likelihood$, i.e.\ the probability of obtaining the observed data $D$ given a particular set of parameters 
$\theta$ for model ${\cal M}$ and prior information ${\cal I}$. $p(\theta|{\cal M}, {\cal I})$ is the $prior$ probability for the parameters $\theta$ of model ${\cal M}$ and $p(D|{\cal M}, {\cal I})$ is the $evidence$, i.e.\ the probability of obtaining the observed data given that model ${\cal M}$ is true. The evidence is defined as
\begin{equation}\label{eq:evidence}
p(D|\mathcal{M,I}) = \int_{\bf{\Theta}}p(\theta|\mathcal{M},{\cal I})p(D|\mathcal{M},\theta,{\cal I})d\theta .
\end{equation}
In other words the evidence is calculated by marginalising the likelihood over the space of the model parameters. It is then obvious that the calculated evidence depends on
the choice of prior range for the parameters $\theta$.

Suppose there are several competing models $\mathcal{M}_i$ that can explain the data. The probability for each model given the observed data $D$ is $p(\mathcal{M}_i|D,{\cal I})$. Applying Bayes' theorem, we obtain
\begin{equation}\label{eq:modpost}
        p(\mathcal{M}_i|D,{\cal I})=\frac{p(D|\mathcal{M}_i,{\cal I})p(\mathcal{M}_i|I)}{p(D|{\cal I})}.
\end{equation}
The \emph{odds ratio} for two different models $\mathcal{M}_i$ and $\mathcal{M}_j$ can then be constructed as
\begin{equation}\label{eq:oddrat}
O_{ij}=\frac{p(\mathcal{M}_i|D,{\cal I})}{p(\mathcal{M}_j|D,{\cal I})}=\frac{p(D|\mathcal{M}_i,{\cal I})p(\mathcal{M}_i|{\cal I})}{p(D|\mathcal{M}_j,{\cal I})p(\mathcal{M}_j|{\cal I})}.
\end{equation}

Normally, in the absence of further information, the prior probabilities for the two models are taken to be equal.  In this case the odds ratio reduces to the ratio of the marginalised likelihoods, or evidences, which is known as the \emph{Bayes Factor}. Usually, a Bayes Factor of 10 is already strong enough to favour one model over the other, while a Bayes Factor of 100 will be regarded as decisive \citep{Jeffreys:1961}. 

The calculation of the evidence can be a time-consuming process for moderate-to-large dimensional parameter spaces.
However the method of nested sampling \citep{Skilling2006} makes
it practicable to calculate evidence efficiently. In this work, we follow closely the method proposed in
\cite{Veitch2010} to calculate the Bayes Factor using nested sampling. 

\section{Results}\label{sec:result}
In our analysis, the two models under consideration only differ in the sign of the frequency change for the second event.
In order to avoid undue influence of the prior range on the Bayes Factor, we assign identical prior ranges to all common parameters in both models. 

\subsection{Setting the Priors}
In table 1 of \cite{Archibald2013}, the parameters are given to be $\nu_0=0.143285110\pm(4\times10^{-9})  {\rm Hz}$, $\dot{\nu}=-9.80\pm0.09\times10^{-15} {\rm  Hz}~{\rm s}^{-1}$. The Epoch (MJD)  is  the time $t_0$ when $\nu(t_0) = \nu_0$. Since the magnetar has been observed for a long time \citep[e.g.][]{2009PhDT........13D,Kaspi2003}, and the spin before the anti-glitch is not of interest, we fix those parameters to be constants.

In both models, there are two independent (anti-)glitch events and for each event there are 4 parameters required to describe it: its epoch $t$ and the changes in the frequency $\Delta \nu$, its first derivative $\Delta \dot{\nu}$ and the phase $\Delta \phi_g$ caused by the event. Thus, in total there are eight parameters for each model. In order that the two models should have a common parametrisation we suppose that in the second model, after the second event (which is a normal glitch in this model), the frequency becomes $\nu_g=\nu-\Delta\nu$ while in the first model, after the second event (which is an anti-glitch) the frequency becomes $\nu_g=\nu+\Delta\nu$. In this way $\Delta\nu$ is a positive parameter for the second event in both models. With this design, the two models can have exactly the same priors, thus minimising the influence of the choice of prior on the final value of the Bayes Factor.

For the epoch of the first anti-glitch event there is an obvious change in flux between the 19th and 20th observation; hence we set the prior for the epoch to be flat between these two data points. For the priors on other parameters we make use of their estimated values $\theta$, together with their uncertainties $\sigma_\theta$, as reported in \citet{Archibald2013}. Specifically we adopt a conservative, uniform prior of width equal to $2n$ times the uncertainty for each parameter -- where we will adopt different values of $n$ in order to explore the robustness of our results to the choice of prior, i.e.\ to check that our prior boundaries contain the vast bulk of the likelihood. 

Thus for each parameter (and where $M_i$ refers to model $i$) the lower boundary of the uniform prior is set to be $\min(\theta_{M_1}-n\sigma_{\theta; M_1},\theta_{M_2}-n\sigma_{\theta; M_2})$, while the upper boundary is set to be $\max(\theta_{M_1}+n\sigma_{\theta; M_1},\theta_{M_2}+n\sigma_{\theta; M_2})$.
Note however, that since $\Delta\nu$ for the second event is always positive, its lower limit is set to be $\min(0,{\Delta\nu}_{M_1}-n\sigma_{\Delta\nu; M_1}, {\Delta\nu}_{M_2}-n\sigma_{\Delta\nu; M_2})$.

\subsection{Comparing the Models}

To calculate the evidence, a nested sampling code was applied with a stopping criterion set to equal $e^{-5}$ -- i.e.\ when new live points made an additional contribution to the evidence that was smaller than a fraction $e^{-5}$ of the total, the nested sampling code was stopped. The value of $n$ used for setting our priors was initially taken to be 10 -- i.e. far beyond the $5\sigma$ region. The two models were found to have evidence values of $\sim e^{-33}$ and $\sim e^{-29}$ respectively, which yields a Bayes Factor of $42.5\pm3.4$ in favour of the successive anti-glitch model over the anti/normal glitch pair model.
According to the definition of \cite{Jeffreys:1961}, a Bayes Factor larger than 40 is already very strong evidence in favour of the successive anti-glitch model.

\begin{figure}[h]
\centering
\includegraphics[width=0.5\textwidth]{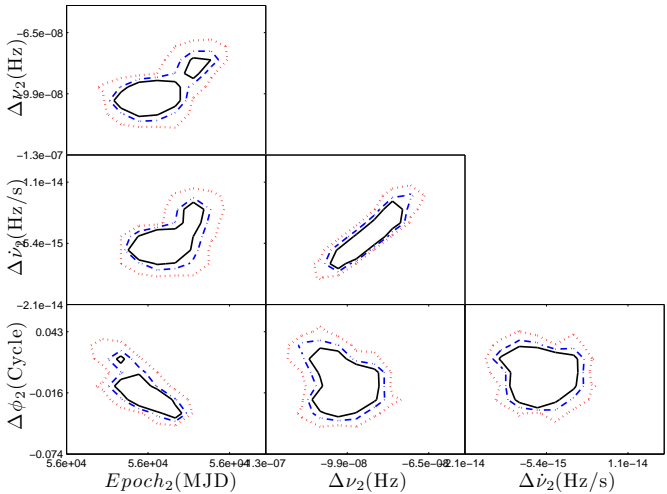}
\caption{Parameter posterior contours for the second anti-glitch event, showing $68.3\%$ (solid), $95.5\%$ (dash-dotted) and $99.7\%$ (dotted) credible intervals, based on $\sim500$ points resampled from the nested sampling samples.}
\label{fig:PE}
\end{figure}

For each model the posterior distribution of the model parameters was resampled appropriately from our nested sampling results. In figure \ref{fig:PE} we show posteriors for the parameters of the second event in the double anti-glitch model. The contour lines correspond to $68.3\%, 95.5\%$ and $99.7\%$ credible intervals. The maximum posterior corresponds to the following best-fitting parameter values: epoch = MJD $56088.4$; $\Delta\nu = -8.2 \times 10^{-8}$ Hz; $\Delta\dot{\nu} = 5.2 \times 10^{-15}$ Hz/s; phase change $= -0.012$ cycles.

\subsection{Robustness Check}

We tested the robustness of our results by changing the width of our uniform priors and re-running the nested sampling
analysis. Table \ref{tab:BF} shows the mean Bayes Factor obtained as $n$ is changed from $10$ to $5$ to $3$, given $10$
independent nested sampling runs for each $n$. We see that the Bayes Factor fluctuates around a value of $\sim45$, but
in all cases our conclusions are consistent.


\begin{figure}[h]
\centering
\includegraphics[width=0.5\textwidth]{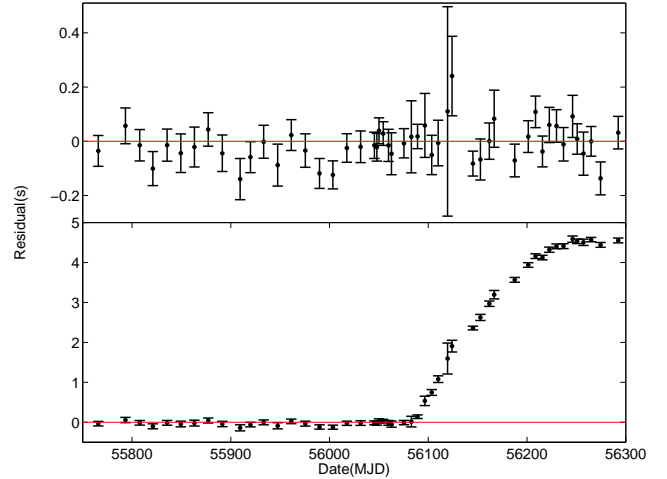}
\caption{Upper panel: Timing residuals of the observed data for two successive anti-glitch events, with best fit parameters as determined by our analysis. Lower panel: Timing residuals of the observed data for only one anti-glitch event. Clearly, two anti-glitch events better explain the data than having a single anti-glitch.}
\label{fig:res1}
\end{figure}

The timing residuals, after subtracting the best fitting double anti-glitch model, are shown in the upper panel of figure \ref{fig:res1}. This consistency confirms that two anti-glitch events can explain the observed data well.

So far, there have been some physical models proposed in order to explain the putative anti-glitch event. Some authors \citep[e.g.][]{Tong2013a,Lyutikov2013} have suggested that the second timing event is not consistent with these physical models, and moreover have questioned whether the observational evidence for the second event is strong enough in the first place. 
However, as shown in the lower panel of figure~\ref{fig:res1}, if we consider only the first event the timing residual will quickly diverge away from zero thereafter, thus showing strong support for the existence of a second timing event. Note that the timing residuals for an anti/normal glitch pair model are also similar to the upper panel in figure~\ref{fig:res1}, further supporting the case for a second timing event (but emphasising that to distinguish between an anti/normal glitch pair and an anti-glitch pair is less straightforward).

We also applied model selection to the case of two timing events versus 1 anti-glitch, and the Bayes Factor was found to be $e^{208}$ -- i.e. overwhelmingly favouring the 2 events scenario. This result demonstrates how the Bayes Factor can favour a more complicated model, notwithstanding that it may require additional parameters, when the data are of sufficient quality and a simpler model cannot give a satisfactory fit.

Finally a batch of simulated glitch-free residual data was also generated, with each point drawn from a Gaussian distribution using means and standard deviations from TEMPO2 fits and TOA errors. Nested sampling was applied to this simulated data, and the Bayes factor was computed for the comparison of a successive anti-glitch model and an anti/normal glitch pair model. We calculated 15 Bayes Factor ratios based on 15 realisations of fake anti/normal glitch-free data. We found that the Bayes Factors fluctuated around unity, i.e. $\exp(0.17\pm0.33)$, showing that intrinsic randomness in glitch-free data will not cause a preference of one model over the other.

\begin{table}
\centering
\caption{Estimated Bayes Factors with uncertainties, obtained from different prior ranges on the model parameters, represented by different $n$ values (for the definition of $n$ see main text). A Bayes Factor of around 45 is obtained in each case, indicating consistently strong evidence favouring a successive anti-glitch scenario over an anti/normal glitch pair.}
  \label{tab:BF}
  \begin{tabular}{@{}cccccc@{}}
  \hline

   $n$ value & 10  & 5 & 3 \\
  \hline

Bayes Factor& $42.4\pm3.4$  	 &  $43.5\pm 3.1$	&  $48.6\pm3.4$\\

\hline
\end{tabular}
\end{table}

\section{Discussion and Conclusions}\label{sec:CD}


We have shown that a model with two successive anti-glitches better explains the observed pulsar data presented in~\citet{Archibald2013} when compared with an anti/normal glitch pair model. Our analysis was robust against variations in the prior ranges, with a Bayes Factor consistently larger than 40 in favour of two anti-glitches. Meanwhile, the Bayes Factor between two events and one event is very large ($e^{208}$), showing conclusively that the two events scenario if favoured over one event.

Prior to the discovery of an anti-glitch there were already several published papers presenting mechanisms that could cause enhanced spin-down, while after its discovery a number of further mechanisms have been proposed seeking to explain its physical origin. Roughly speaking, we can divide the proposed mechanisms into three groups: \emph{internal , accretion} and \emph{magnetoshperical}. 

The internal mechanism is related to that causing glitches in normal pulsars, which can often be satisfactorily explained by the coupling of the crust with the inner faster-rotating superfluid \citep{1996ApJ...457..844L}, where for a normal pulsar the superfluid interior could not spin slower than the crust. 
However, as the observed object 1E 2259+586 is a magnetar, where the dominant source of free energy is magnetism instead of rotation, the spin evolution could be vastly different from that of normal pulsars. \cite{1995MNRAS.275..255T} suggested that a magnetar could drive differential rotation, which allows a lag in the rotation of the superfluid interior. A sudden rearrangement of the inner structure could induce the interior and crust to corotate again, which would be observed as a sudden spin-down, or anti-glitch \citep{Archibald2013}.
Another possible explanation for the faster-rotating crust might be the twist of a crust patch. As the superfluid vortex is pinned to the crust, a plastic deformation for such a patch will lead to a slower rotating superfluid. A rapid twist would correspond to a conventional spin-up glitch, similar to a normal pulsar counter-part.  However, while a gradual twist would have little effect on the secular spin evolution, a rapid unpinning of
the associated vortices would give a sudden spin-down, or anti-glitch \citep{Duncan2013,Thompson2000} 
 
Accretion mechanisms suggest that the anti-glitch is caused by the accretion of retrograde material from either a Keplerian ring \citep{Ouyed2013} or from an asteroid \citep{Huang2013}. Besides retrograde accretion, \citet{Katz2013} also proposes an enhanced propeller effect to explain the anti-glitch. Although most accretion models are able to explain both events during the observation, either with or without being accompanied by radiation, this mechanism does not fit the model of magnetars, which has already been supported by many observations.~\citep[e.g.][]{Kaspi2003}.

Magnetospheric models \citep[e.g.][]{Tong2013a,Lyutikov2013,Thompson2000} explain the observed anti-glitch with either an enhanced particle wind or a twisting of the magnetic field lines. Although these models fit the observational data for magnetars, most magnetospherical explanations are accompanied with strong radiation and/or a change in pulse profile -- neither of which were observed during the second timing event for 1E 2259+586. 
The magnetospheric mechanism is not favoured since figure~\ref{fig:res1} 
shows that our analysis strongly favours the existence of the second event.

Among these three mechanisms, our analysis shows that the internal mechanism is most favoured.
We note that a satisfactory model should be able to explain the two successive anti-glitches that happened within a relatively short period. If the sudden unpinning of the quantum vortex due to the twist of crust patch is responsible for the anti-glitch, for example, then the gradual plastic deformation of the crust patch should be able to accumulate enough angular momentum within a timescale of several months. If the two anti-glitches are caused by the same mechanism, then the observations may put some constraints on that mechanism. Enhanced radiation, pulse profile changes and enhanced spin-down were observed for the first event while none of these phenomena was observed for the second. 
Future observations of similar phenomena with higher timing accuracy and sampling frequency will be helpful in order to more fully understand the mechanism responsible. 

\section{Acknowledgments}
The authors would like to thank Robert Archibald, Victoria Kaspi and Neil Gehrels for helping with the data and discussions. We are also grateful for further useful discussions with Jonathan Katz, Yongfeng Huang and Hao Tong. Additionally, we give special thanks to Ian Jones for his very helpful feedback on our article. The authors would also like to thank the anonymous reviewers for constructive comments that helped in improving the quality of this manuscript. The authors gratefully acknowledge the support of the Science and Technology Facilities Council of the United Kingdom and the Scottish Universities Physics Alliance. Y.-M. Hu is supported by the China Scholarship Council.

\end{document}